\documentstyle[preprint,aps,epsf]{revtex}
\newif\iftightenlines\tightenlinesfalse
\tightenlines\tightenlinestrue
\begin{document}
%
\def\pT{p_T^{\phantom{7}}}
\def\MW{M_W^{\phantom{7}}}
\def\ET{E_T^{\phantom{7}}}
\def\bh{\bar h}
\def\lm{\,{\rm lm}}
\def\tG{\tilde G}
\def\lo{\lambda_1}                                              
\def\lt{\lambda_2}
\def\ETC{E_T^c}
\def\pslt{p\llap/_T}
\def\eslt{E\llap/_T}
\def\etmiss{E\llap/_T}
\def\eslt{E\llap/_T}
\def\to{\rightarrow}
\def\Re{{\cal R \mskip-4mu \lower.1ex \hbox{\it e}}\,}
\def\Im{{\cal I \mskip-5mu \lower.1ex \hbox{\it m}}\,}
\def\SU{SU(2)$\times$U(1)$_Y$}
\def\te{\tilde e}
\def \tlam{\tilde{\lambda}}
\def\tl{\tilde l}
\def\tb{\tilde b}
\def\tst{\tilde t}
\def\tt{\tilde t}
\def\ttau{\tilde \tau}
\def\tmu{\tilde \mu}
\def\tg{\tilde g}
\def\tga{\tilde \gamma}
\def\tnu{\tilde\nu}
\def\tell{\tilde\ell}
\def\tq{\tilde q}
\def\tw{\widetilde W}
\def\tz{\widetilde Z}
\def\cmsec{{\rm cm^{-2}s^{-1}}}
\def\fb{{\rm fb}}
\def\sgn{\mathop{\rm sgn}}
\def\mhf{m_{\frac{1}{2}}}

\hyphenation{mssm}
\def\ds{\displaystyle}
\def\ts{${\strut\atop\strut}$}
%
%
\preprint{\vbox{\hbox{FSU-HEP-000914}
                \hbox{UH-511-973-00}}}

%
\title{Analysis of Long-Lived Slepton NLSP in GMSB model at Linear
Collider} 
\author{Pedro G.\ Mercadante$^1$, J.\ Kenichi Mizukoshi$^2$, and Hitoshi 
Yamamoto$^{2}$}
\address{
$^1$Department of Physics,
Florida State University,
Tallahassee, FL 32306, USA
}
\address{
$^2$Department of Physics and Astronomy,
University of Hawaii,
Honolulu, HI 96822, USA
}
\date{\today}
\maketitle
\begin{abstract}
We performed an analysis on the detection of a long-lived slepton at a
linear collider with $\sqrt{s}=500$ GeV. In GMSB models a long-lived NLSP
is predicted for large value of the supersymmetry breaking scale 
$\sqrt{F}$.
Furthermore in a large portion of the
parameter space this particle is a stau. Such heavy charged particles
will leave a track in the tracking volume and hit the muonic detector. 
In order to
disentangle this signal from the muon background, we explore 
kinematics and particle identification tools: time of flight device, dE/dX
and Cerenkov devices. We show that a linear collider will be able to 
detect long-lived staus with masses up to the kinematical limit of the 
machine. We also present our estimation of the sensitivity to the stau 
lifetime.

\end{abstract}

\medskip

\section{Introduction}
In many supersymmetric models charged long-lived particle can exist. In models
with $R-$parity violation, this happens when the lightest supersymmetric 
particle (LSP) is a slepton and the $R-$parity violation term is small 
\cite{krasnikov}. Another class consist of models where the LSP is not the
usual (like in mSUGRA) $U(1)_Y$ gaugino but rather a $SU(2)$ gaugino; 
in this case 
the next-to-lightest supersymmetric particle (NLSP) is the lightest chargino, 
almost degenerate in mass with the neutralino LSP, and  will decay after 
traveling centimeters  or meters  into the LSP plus a soft 
lepton or pion \cite{su}. This type of model arises rather naturally in 
the anomaly mediated SUSY breaking scenario \cite{anomaly}, however in this 
scenario the mass splitting is typically larger then the pion mass, leading 
to decay length of the order of $c\tau\sim 10$ cm at most. Another class of 
models where long-lived charged particles occur is in gauge mediated 
supersymmetry breaking (GMSB) models~\cite{early,dine}. Depending on 
the value assumed for the SUSY breaking scale, $\sqrt{F}$, the gravitino
is the LSP. Moreover, due to the weak coupling of the gravitino to the 
SUSY fields, the NLSP could in principle be a long-lived massive particle. 
Throughout this paper we will concentrate on the detection of long-lived 
charged particles in the  GMSB framework. The techniques presented, 
however, are applicable to general non-strongly interacting 
heavy long-lived charged particles \footnote{In the Ref.~\cite{drees},
the authors raise the issue of hadronization effects for color-triplet
particles. Some of them will hadronize into neutral exotic 
mesons. Besides that, the inelastic hadronic reactions can change the 
charge of the mesons.}.

In GMSB an intermediate sector is responsible for communicating
Supersymmetry breaking to the MSSM sector. In these models the gravitino 
mass is given by,

\begin{eqnarray}
M_{\tilde{G}}=\frac{F}{\sqrt{3}M_{\rm pl}}=\left(\frac{\sqrt{F}}{100 
\;\mbox{TeV}}\right)^2 2.37 \; \mbox{eV}
\end{eqnarray}
where  $M_{\rm pl}$ is the reduced Planck mass. For 
the $\sqrt{F} \sim$ 100 TeV (to be compaired with $10^{11}$ GeV, a typical 
mSUGRA value of SUSY breaking scale), $M_{\tilde{G}} \sim$ eV, making the 
phenomenology quite interesting. 

 Moreover, because
of the very weak interaction of the gravitino, all supersymmetric
particles will decay into the NLSP and this will decay into the gravitino 
(which will escape
detection) and its standard model partner. Thus, the nature of the NLSP and
its decay length
will play a fundamental role in the phenomenology of the
model.
Because soft terms in GMSB models are generated by the gauge couplings,
the NLSP is usually either a neutralino or a stau. Many phenomenological 
studies have been made for Tevatron and
LEP in the context of this model~\cite{dim,gunion,rg}.
More recently some studies for LHC and Run II at Tevatron also have been 
addressed~\cite{bmtw,ambro2,tevrun2}.
For the next linear collider (NLC) the case for a  
neutralino NLSP,
both long- and short-lived, was considered by Ambrosanio~\cite{ambro}
and the case for a
stau  NLSP with prompt decay was considered by Kanaya~\cite{work}.
 
In this work we improve over our previous analyses~\cite{work,my} 
in the search for a long-lived stau at NLC, presenting 
here our estimations for a lifetime measurement. 
In the GMSB the  NLSP lifetime is given by,
\begin{eqnarray}
c\tau=16\pi\frac{F^2}{M^5_{\mbox{\tiny NLSP}}}\sim 
\left(\frac{\sqrt{F}}{10^7 \;\mbox{GeV}}\right)^4\left(\frac{100 \;\mbox{GeV}}
{M_{\mbox{\tiny NLSP}}}\right)^5 
 10 \; \mbox{km}.
\label{eq:ctau}
\end{eqnarray}

From this expression, one can see that  the  measurement of the NLSP mass and 
lifetime will determine the value  of the fundamental SUSY breaking 
scale $\sqrt{F}$. From perturbative 
arguments it is possible to set a lower limit in $\sqrt{F}$ for the complete
set of parameters in the GMSB framework. Nevertheless, for the range of stau
masses under consideration, it is always possible to find parameters which 
gives a $c\tau \sim \mu$m. On the other hand, there is no solid theoretical 
argument for an upper limit on $\sqrt{F}$. However,  a LSP gravitino with 
mass higher than few KeV is disfavored in some cosmological 
scenarios for over-closing the Universe \cite{cosmo}.  This translates to 
roughly $\sqrt{F} \lesssim 3000$ TeV.

The rest of this  paper is organized as follows. In section II we describe 
our simulation using the program ISAJET \cite{isajet} to generate 
stau pairs with the effects of a time of flight device, dE/dX and a 
Cerenkov device to detect a heavy particle. In section III we use the 
stau pair production process to extract limits on the lifetime of 
stau which leads to limits on the  SUSY breaking scale  in a GMSB model. 
In section IV we present our conclusions.

\section{Selection criteria for stau pair production}

Stau pair production at a linear collider provides a clean search
environment. Furthermore, the production cross section is
model-independent,  depending  only on the mass of the
staus and on the mixing angle between the left- and right-handed 
superpartners.       
In Fig.~\ref{fig:1} we show the pair production cross section
(normalized to $\sigma_{\mu \mu}=450$ fb) for the left- and right-handed
states as a function of stau mass. We can see that
the stau pair cross section is smaller than $\sigma_{\mu \mu}$
 due to its scalar nature and it rapidly drops when we approach the 
kinematical limits of the
accelerator. Nevertheless, we note that for masses around $240$
GeV we still have cross section of ${\cal{O}} (10)$ fb, which  should be
observable provided that the background is manageable.

We will first describe the selection criteria without particle identification
which will be used in later sections to extend the range of sensitivity
to the full beam energy. The signal we are looking for is a back-to-back 
tracks with
corresponding hits in the muon chamber. With this requirement
tracks from $\pi, K, p$ and $e$ are removed.  To
reject the two photon process of 
$\gamma \gamma \rightarrow \mu^+ \mu^- $,
we note that the $\mu^+ \mu^-$ pair in this process tends to have a low 
invariant mass
and be boosted along the beam pipe. Thus, we require the following cuts:
\begin{enumerate}
\item{$\cos{\theta} < 0.8$, to guarantee good track quality;}
\item{$|p| > 0.5 E_{\mbox{beam}}$ and} 
\item{$|p^{\mbox{tot}}_z| < 0.25  E_{\mbox{beam}}$}.
\end{enumerate}
After these cuts, the two photon initiated muon pair
production is estimated to be 1.4 fb. We are then left with muon pair
production $e^+ e^- \rightarrow \mu^+ \mu^-$
as the main source of background.

In order to reduce the muon pair background we shall explore the heavy mass
of the staus.  In a $e^+ e^-$ collider the energy in the
center-of-mass is fixed, so in a pair production process the energy of
the final particles is also known. It is well known, however, that
in a high energy linear collider beamsstrahlung and initial state
radiation  effects become
important and the effective energy of the reaction is not fixed
but presents a spectrum. In ISAJET, these effects have been 
implemented using the  parameterization described in  Ref.~ \cite{psim}, 
where the
radiation spectrum is well approximated by one
photon emission from one of the initial state $e^{\pm}$. Assuming also
that the direction of the emission is along the beam line, 
the mass estimate for each track in the laboratory frame is given by:

\begin{eqnarray}           
\label{eq:m2}
M^2&=&\left( \frac{\sqrt{\hat{s}}}{2\gamma} + \beta p_z
 \right)^2 - |p|^2 \;,\\
\hat{s}&=&s(1-|\Delta|) \;,\\
\beta&=&\Delta/(2-|\Delta|) \;,
\end{eqnarray}
where $\Delta=p_z^{\mbox{tot}}/E_{\mbox{beam}}$ is the net momentum in the
beam line direction, $\beta$ is the boost parameter and $\sqrt{\hat{s}}$ is
the center-of-mass energy of the two tracks.

In Fig.~\ref{fig:2} is shown the cross section distribution as a function of
$M^2$ which is estimated according to Eq.~(\ref{eq:m2}). In this plot we 
can see the muon
distribution peaking at zero mass with a tail from beamsstrahlung. We
also see the distribution of staus  production for several values of 
masses. The momentum resolution is taken to be $\delta p_T/p_T=5
\times 10^{-5}
p_T$ (GeV). Based in this plot we use the following cut,

\begin{enumerate}
\setcounter{enumi}{3}
\item {$|M^2-M^2_{\tilde{\tau}}| < 3000 \;\; \mbox{GeV}^2$},
\end{enumerate}
where $M_{\tilde{\tau}}$ is a
variable parameter in the search. The resulting efficiency after this cut
is shown as  the dotted line of Fig.~\ref{fig:3}a. This is our basic 
strategy to reduce 
the muon background. To further improve the sensitivity, we study particle 
identification.

\subsection{Time of Flight}

A time of flight (TOF) device can be used to identify heavy tracks. In our
study we considered a linear collider with $1.4$ ns of bunch
separation. In the large detector  scenario ($r=2$ m) the mean time of
flight for a massless ($\beta=1$) particle is around $6.7$ ns. Assuming
that we do not  know which bunch crossing a given event is
coming from, every 1.4 ns of TOF delay is consistent with the
massless muon.  We
simulated the effect of a $50$ ps error in the time of flight
measurement, applying the following cut,
\begin{enumerate}
\setcounter{enumi}{4}
\item{$\Delta t > 0.13$ ns,}
\end{enumerate}
where $\Delta t$ is the time of flight difference between a $\beta=1$ and a
massive particle, modulo $1.4$ ns.

This cut corresponds to about
$2.5 \sigma$, so that about $1\%$ of the muon background  are kept.  
Applying this
cut to both tracks we can relax cuts 2 and 3 which extends the mass
range to the full beam energy. The
efficiency of this cut as a function of mass is shown in 
Fig.~\ref{fig:3}a as the dashed line. The corresponding sensitivity is
shown in Fig.~\ref{fig:3}b.
We see that for some values of the mass we have lower 
efficiency when the time delay with respect to $\beta = 1$ case becomes
equal to the bunch spacing. However, the efficiency is never below 0.3 
because the actual path
that the particle goes through depends on the angle (the detector has 
cylindrical, but not spherical, symmetry): for each angle we have a different
travel length.

\subsection{dE/dX}

When a charged particle goes through the detector it deposits
energy by ionization. The amount of energy deposited, dE/dX, is a function of 
$\beta \gamma$ of the particle~\cite{pdg}. We assumed that the charged particle
goes through argonne and dE/dX has 5\% resolution, which is a realistic 
value for a TPC tracking chamber. To remove muons, we 
use the following cut,
\begin{enumerate}
\setcounter{enumi}{5}
{\item  $\frac{dE/dX - dE/dX(\rm{muon})}{\sigma(dE/dX)} > 3 $}.
\end{enumerate}
The resulting efficiency is shown as
the solid line in Fig.~\ref{fig:3}a. We note a blind spot for masses around
$150$ GeV. We also note that for lower value of $\beta \gamma$ (for masses
below $150$ GeV) dE/dX will give a unique value for $\beta \gamma$, thus 
knowing the momentum we can get the stau mass. As the figure shows, we 
can apply cut 6 to cover the high mass range up to
the beam energy where the cuts 2 and 3 are not met.

In Fig.~\ref{fig:3}b we present, for each strategy, the minimum cross section
(beforee cuts) that will  be visible
at a $\sqrt{s}=500$ GeV linear collider with $L=50\;\; {\mbox
{fb}}^{-1}$.  Our criteria is based on a 3 sigma significance; namely,
$S\ge 3$ with
$S=\epsilon \sigma \sqrt{L/bg}$, where $\sigma$ is the
 signal cross
section (before cuts), $\epsilon$ is the efficiency to pass the cuts
and $bg$ is the
expected  background cross section after cuts. We require a minimum of
5 signal events after cuts.

A Cerenkov device can be used to measure $\beta \gamma$. With a device
similar to the BABAR detector~\cite{babar} it is
possible to reject particles with $\beta \gamma > 8$. 
Cerenkov devices, however, in general have a large impact on the
detector design due to its requirements on space and photon 
detection. If these requirements are met, a Cerenkov device can replace 
TOF or dE/dX discussed above.

A  comment on the nature of our results is in order. We have presented a
strategy based only on the pair production mechanism where particle 
identification had a relatively minor role to play. Nevertheless, in the 
models under consideration it is likely that others supersymmetric particles 
will be produced and end up in stau; in such decay chains the use of time of 
flight, dE/dX and Cerenkov devices would play a critical role in identifying 
staus \footnote{The kinematical
distribution, as proposed here, would be of no use in a process other
than pair production.}. We also note that the results presented so far in 
Fig.~\ref{fig:3} are rather independent of model; it depends only in the
pair production mechanism that will exist for any non-colored 
charged long-lived particle, provided that the cross section is not too small. 
The efficiency to pass the cuts depends only on the mass of the 
particle \footnote{There is a model dependent 
part in the angular distribution of the cross section that will have an effect
 in cut 1. But we note that the effect of this cut is small. One can also
read Fig.~\ref{fig:3} as the minimum cross section in the central region that 
will be observed at the NLC.}, 
so that Fig.~\ref{fig:3} can be read in a model independent way: we see 
the minimal 
cross section that would lead to an observable signal as a function of the
mass of a long-lived charged particle, regardless of the nature of this 
particle.

\section{stau decay length}

Up to this point we supposed that the stau does not decay within
the detector; however, as mentioned
earlier, the lifetime can be viewed as a free parameter of the
model. We will now discuss a simple measurement of the stau lifetime
using the mode $e^+ e^- \rightarrow \tilde{\tau}^+  \tilde{\tau}^-$.
The case of a short-lived stau is to be studied elsewhere.

In order to ensure good measurement of momentum and dE/dX 
 we require that each event should have one central track longer than 1
meter, using the mass cut $4$ to select tracks consistent with a heavy
particle. We
believe that such events will be essentially background free: a well
reconstructed track, consistent kinematically with a
heavy particle, regardless of the decay pattern.

One way to get the lifetime is to consider the number of events that
decay before and after a certain length. With this method the error in
the lifetime is given by:

\begin{eqnarray}
\label{eq:6}
\frac{\sigma_{c\tau}}{c\tau}&=&\frac{1}{\sqrt{N}}
\frac{\sqrt{R}}{\log{(1+R)}} \\
R&=&\frac{N_1}{N_{\inf}}
\end{eqnarray}
where $N_1$ is the number of particles that decays between distance
$l_1$ 
and $l_2$, $N_{\inf}$ is the number of particles that
decays after
distance $l_2$ and $N=N_1 + N_{\inf}$ is the total number of particle
that  decay after
distance $l_1$. In practice, $l_1=1\rm{m}$, and $l_2$ is chosen such that
the statistical power is maximized, which is found to occur at $R=3.9$,
but not exceeding the outer radius of the tracking volume. When the 
lifetime is short enough the optimum value of $l_2$ occurs within the
tracking volume, and the error [Eq.~(\ref{eq:6})] is given by
$\frac{1.24}{\sqrt{N}}$. On the other hand, a large lifetime may lead to the
optimum $l_2$ outside of the tracking volume in which case the 
error is approximately $\frac{1}{\sqrt{N_1}}$. 
In both cases the relevant quantity to get a
precise measurement of the lifetime is the number
of particles that decay inside the detector.

In Fig.~\ref{fig:4}, the solid lines represent the  contour plot
of  the  lifetime resolution as a function of the stau mass and 
lifetime where minimum track length ($l_1$) of 1 m is required.
In many models, stau can be identified without using its own track. 
In order to take these cases into account, we also show the 
lifetime resolution for $l_1 = 1$ cm (the dashed lines). 
We see from this plot that in the optimistic case we would
be able to measure the lifetime with a $10 \%$ precision as long as $c\tau$
is lower (bigger) than $\sim$ 40 m ($\sim 0.01$ m), for masses up 
to $200$ GeV.    
A comparison with  the CERN LHC is in order \cite{ambro2}. In this
study Ambrosanio {\it et al.} show the capability of the CERN LHC to measure 
the stau lifetime
in several models within the GMSB context. Their numbers are parameter
space 
dependent as all possible reactions (that ended in staus at the end of the 
decay chain) are used to extract the stau lifetime. It is of fundamental 
importance for the LHC to consider all possible reaction in order to get 
enough statistics. Their analysis shows that for models that gives 
$M_{stau} \sim 100$ GeV they are able to get $10 \%$ precision 
(in a somewhat optimistic case) for $ 1<c\tau (\mbox{m}) <50$.
 Our analysis involves only the stau pair production,
being essentially model independent \footnote{We note that we used an almost
pure right-handed stau that indeed gives a somewhat lower cross section 
than a left-handed stau.}. For $M_{stau} \sim 100$ GeV we are able to get
$ 0.0013<c\tau (\mbox{m}) <50$  ($ 0.13<c\tau (\mbox{m}) <25$)  for the 
minimal track length of 1 cm (1 m) requirement. We also note that we 
are using a somewhat conservative  integral luminosity of $50$ fb$^{-1}$. 

Also in Fig.~\ref{fig:4} is shown in dotted lines a $90 \%$ confidence 
level upper limit for $c\tau$,  assuming  that stau decays outside
the detector. From Eq.~({\ref{eq:ctau}), we can see that this value
corresponds to a lower bound $\sqrt{F} \sim 1 \times 10^7$ GeV, which
is  slightly above cosmologically preferred  value.

\section{Conclusions}

In many supersymmetry scenarios a charged long-lived particle is 
predicted. For instance, in the GMSB scenario such particle would be a stau 
in a large portion of parameter space. 
We have studied the long-lived stau pair production in a linear collider
at $\sqrt{s}=500$ GeV. The linear collider will be able to study the 
stau pair where no other particles are produced in the event (except for 
beamsstrahlung, etc.), as opposed to the LHC where all supersymmetric
reactions should be taken into account \cite{ambro2,bmtw}, being in 
principle a good place to extract the parameters of the model. 

We presented in a model independent way the minimal cross section for the
pair production of stable non-strongly interacting charged particle 
that will be observed in the NLC,
considering just momentum measurement as well as  particle identification 
devices. When the predicted cross section for the stau pair production in 
the GMSB model is considered it is shown that the NLC will be able to detect
such reactions for stau masses up to $85 \%$ of the beam energy, with just
momentum measurement. Particle identification devices will extend the mass
range to essentially the full beam energy. Moreover, particle identification
devices will provide a sample of events essentially background free. 

We also presented a way to extract the stau lifetime using
the predicted cross section and a luminosity of $50$ fb$^{-1}$. The 
method presented is very straightforward and does not depend in any other
parameter of the model. For the range of masses that is possible to probe
in a $500$ GeV linear collider, the precision obtained is, in general, 
better then it will be possible to have from the LHC and it is mostly
model independent.

\acknowledgments

We thank X.~Tata for valuable discussions and a careful reading of the 
manuscript. This work was supported by the 
U.~S.~ Department of Energy under contract DE-FG02-97ER41022
and DE-FG03-94ER40833. During his stay at the University of Hawaii 
where this work was begun, P.~G.~M.~ was partially supported by 
Funda\c{c}\~ao de Amparo \`a Pesquisa do Estado de S\~ao Paulo (FAPESP).

%

%
%
%
%


\iftightenlines\else\newpage\fi
\iftightenlines\global\firstfigfalse\fi
\def\dofig#1#2{\iftightenlines\epsfxsize=#1\centerline{\epsfbox{#2}}\bigskip\fi}

\begin{figure}
\dofig{5in}{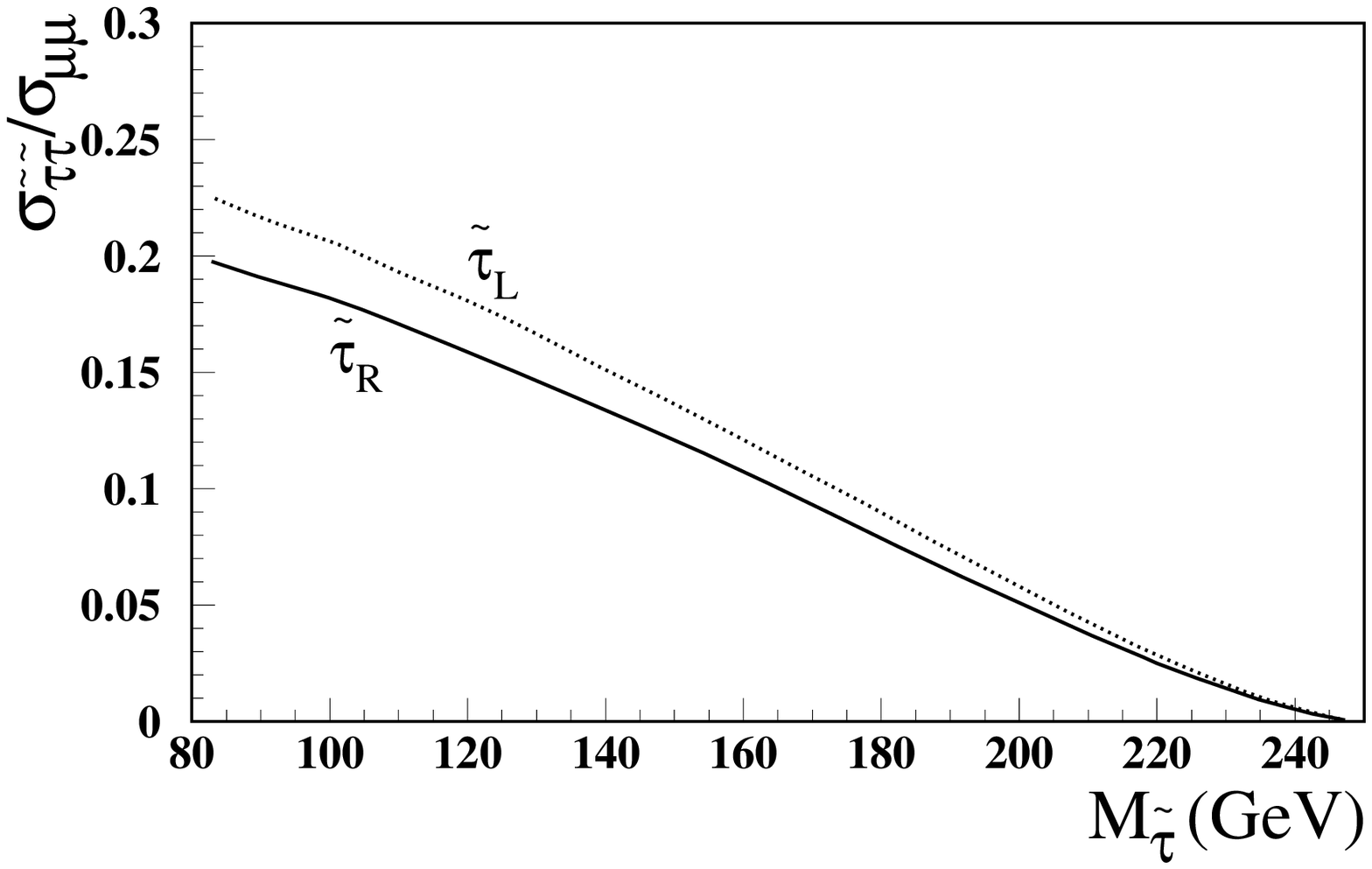}
\caption[]{Stau pair production cross section normalized by muon pair as a 
function of stau mass in $\sqrt{s}=500$ GeV linear collider. The full
(dotted) line is for the right- (left-) handed stau pair.}
\label{fig:1}
\end{figure}

\begin{figure}
\dofig{5in}{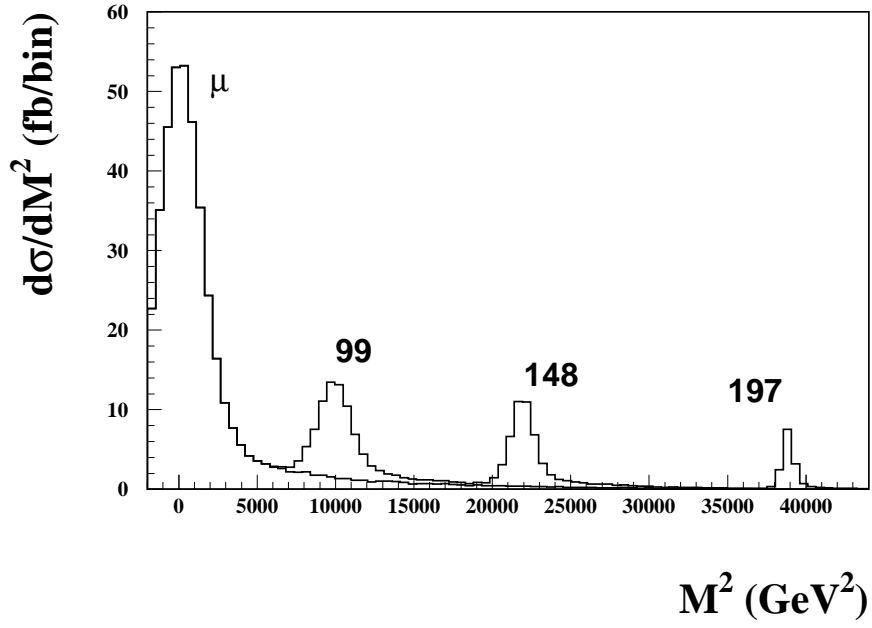}
\caption[]{The mass square distribution for the muon and for 
several values of stau masses.}
\label{fig:2}
\end{figure}

\begin{figure}
\dofig{5in}{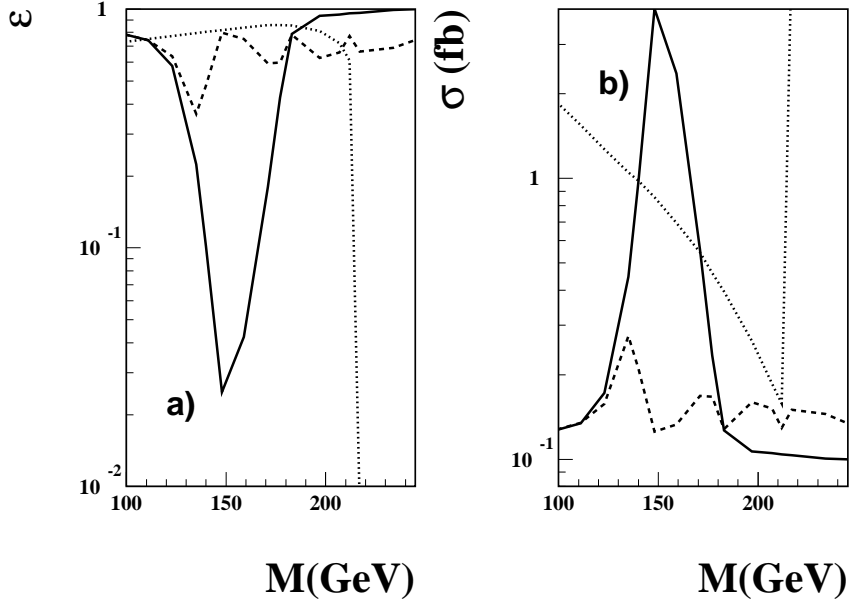}
\caption[]{a) The efficiency for the signal. The dotted line is the
efficiency for just kinematics cuts (cuts 1-4), the dashed line is for the
time of flight cut (cuts 1 and 5) and the solid line is for dE/dX
cuts (cuts 1 and 6). b) The reach in cross section for a 
linear collider with
luminosity of 50 $\mbox{fb}^{-1}$. The dotted line is using just
kinematics, the dashed line is using time of flight and the solid line
is using dE/dX.}
\label{fig:3}
\end{figure}

\begin{figure}
\dofig{5.5in}{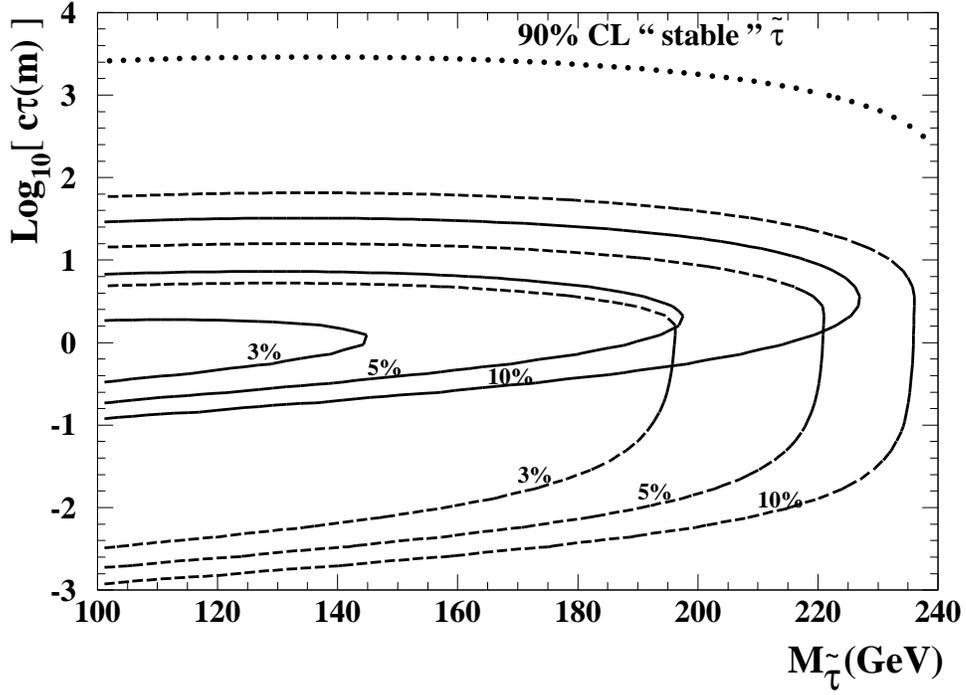}
\caption[]{Contours of constant error $\sigma_{c\tau}$ on the
measurement of $c\tau$. The solid lines stand for the case $l_1=1$ m
and dashed lines are for $l_1 = 0.01$ m. The dotted line on the top of
the figure indicates at 90\% CL the case stau is stable up to the detector.}
\label{fig:4}
\end{figure}

\vfil

\end{document}